\input harvmac
\noblackbox
\Title{UCLA/96/TEP/24}
{Understanding Chiral Anomaly in Coordinate Space $^\star$
\footnote{}{$^\star$ This work was supported in part by the U.S.
Department of Energy, under Contract
DE-AT03-88ER 40384 Mod A006 Task C.}}

\centerline{
Hidenori SONODA$^\dagger$\footnote{}{$^\dagger$
E-mail: sonoda@physics.ucla.edu}}
\bigskip\centerline{\it Department of Physics and Astronomy, UCLA,
Los Angeles, CA 90095-1547, USA}

\vskip 1in
By completing the old
discussion of K.~Wilson, we express the chiral anomaly
in terms of a double integral of a three-point function of
chiral currents over
an arbitrarily small region in the coordinate space.  An integrability
condition provides an important finite local counterterm to the integral.

%\vskip .2in
%\noindent
%PACS numbers: 11.10.-z, 11.10.Gh, 11.30.Rd, 11.40.-q

%\draft
\Date{August 1996}

% begin definitions
\def\vev#1{\left\langle #1 \right\rangle}
\def\ep{\epsilon}
\def\J{\tilde{J}}
\def\tr{\hbox{tr}~}
\def\S{{\cal S}}
% end definitions

All known derivations of the chiral anomaly suggest the short-distance
nature of the effect. In the derivation by Schwinger
\ref\rschwinger{J.~Schwinger, Phys.~Rev.~{\bf 82}(1951)664}, the axial
anomaly appears as an inevitable consequence of the short-distance
singularity between two spinor fields and the vector gauge invariance.
In another derivation
\ref\rabj{S.~L.~Adler, Phys.~Rev.~{\bf 177}(1969)2426\semi
J.~Bell and R.~Jackiw, Nuovo Cimento~{\bf 60A}(1969)47\semi
W.~Bardeen, Phys.~Rev.~{\bf 184}(1969)1848} the anomaly arises from
the ambiguity of a three-point function of chiral currents due to the
linear divergence (in the momentum space) of a loop integral.  If we
regulate the linear divergence by the Pauli--Villars regulator, a
local expression for the anomaly is obtained in the limit of the
infinite regulator mass
\ref\rjackiw{R.~Jackiw, in ``Lectures on Current Algebra
and Its Applications,'' Princeton University Press (Princeton, NJ
1972)}, which is equivalent to the zero distance limit.

The purpose of this little note is to complete and extend the old work
of K.~Wilson
\ref\rwilson{K.~G.~Wilson, Phys.~Rev.~{\bf 179}(1969)1499}
in which the chiral anomaly
was analyzed in the coordinate space.  Though the anomaly was not
evaluated explicitly in ref.~\rwilson,
the origin of the anomaly was correctly identified as the
subtlety in defining a double integral of a three-point function of
chiral currents caused by short distance singularities.  The final
expression we will obtain for the anomaly is valid not only for the
free theory but also for the interacting theories like QCD.  To
facilitate the derivation we will use the euclidean space instead of
the Minkowski space.\foot{The chiral anomaly has been discussed and
computed in the framework of differential renormalization in
coordinate space \ref\rdiff{D.~Z.~Freedman, K.~Johnson, and J.~I.~Lattore,
Nucl.~Phys.~{\bf B371}(1992)353}.  Our approach is more directly
related to ref.~\rwilson.}

We first review consistent and covariant chiral currents.  For
definiteness we consider QCD with $N_f$ massless quarks.  If we couple
a classical SU($N_f$) gauge field $A$ to the right-handed currents,
the vacuum energy $W[A]$ becomes a functional of the
external gauge field.  The
consistent current is defined so that its expectation value gives the
first order variation of the vacuum energy:
\eqn\edelvac{W[A+\delta A] - W[A] = \int d^4 r ~\delta A_\mu^A (r)
\vev{J_\mu^A (r)}_A~.}
The consistent current $J_\mu^A$ satisfies the anomaly equation:
\eqn\econanomaly{D_\mu J_\mu^A \equiv \partial_\mu J_\mu^A
+ f^{ABC} A_\mu^B J_\mu^C =~c~\ep_{\mu\nu\alpha\beta}\partial_\mu
\tr (-i T^A) \left( A_\nu \partial_\alpha A_\beta + {1 \over 2}
A_\nu A_\alpha A_\beta \right)~,}
where $c$ is a constant, $A_\mu \equiv -i T^A A_\mu^A$,
$[T^A, T^B] = i f^{ABC} T^C$,
and the hermitian generators are normalized as $\tr T^A T^B = \delta^{AB}$.
The consistent current $J_\mu^A$ does not transform covariantly under
the flavor gauge transformation.  To make it covariant, we must
add a counterterm:
\eqn\ecov{\J_\mu^A \equiv J_\mu^A + \Delta J_\mu^A~,}
where
\eqn\ecounter{\Delta J_\mu^A \equiv c ~\ep_{\mu\nu\alpha\beta}
\tr (-i T^A) \left( \partial_\nu A_\alpha A_\beta +
A_\nu \partial_\alpha A_\beta + {3 \over 2} A_\nu A_\alpha A_\beta
\right)\quad.}
The covariant current then satisfies
\eqn\ecovanomaly{D_\mu \J_\mu^A
= {3 c \over 4}~\tr (-i T^A) F_{\mu\nu} \tilde{F}_{\mu\nu}~,}
where $F_{\mu\nu}$ is the field strength,
and $\tilde{F}_{\mu\nu} \equiv \epsilon_{\mu\nu\alpha\beta}
F_{\alpha\beta}$.

Our goal is to compute the covariant divergence of the covariant current
$D_\mu \vev{\J_\mu^A}_A$ to second order in the external gauge
field.  In order to calculate this in the coordinate space, we
need a formula that expresses
the first order variation of the n-point function of the
covariant currents.  The formula is given by
\eqn\evar{\eqalign{
&\vev{\J_{\mu_1}^{A_1} (r_1) ~...~ \J_{\mu_n}^{A_n} (r_n)}_{A+\delta A}
- \vev{\J_{\mu_1}^{A_1} (r_1) ~...~ \J_{\mu_n}^{A_n} (r_n)}_A \cr
&\quad = - \int_{|r - r_k| \ge \ep_k} d^4 r~\delta A_\mu^A (r)
\vev{\left(\J_\mu^A (r) - \vev{\J_\mu^A (r)}_A\right)
\J_{\mu_1}^{A_1} (r_1) ~...~ \J_{\mu_n}^{A_n} (r_n)}_A\cr
& \quad \quad - \sum_{k=1}^n ~[\delta A \cdot \S (\ep_k; A(r_k))]_{\mu_k}^{A_k}
\vev{\J_{\mu_1}^{A_1} (r_1) ~...~\widehat{\J_{\mu_k}^{A_k}
(r_k)} ~...~ \J_{\mu_n}^{A_n} (r_n)}_A~.\cr}}
The limit $\ep_k \to 0$ can be taken independently as long as
the points $r_k$ are all different.  The c-number counterterm $\delta A
\cdot \S$ is necessary to subtract the short-distance singularities
contained in the product of two chiral currents.
The possible form of the counterterm
is restricted by covariance, dimensionality, and
CP invariance.  The most general form is given by
\eqn\eS{\eqalign{&[\delta A \cdot \S (\ep;A)]_\mu^A \cr
&= \tr (-i T^A) \Big[
c_1 (\ep) \delta A_\mu
+ c_2 (\ep) [\delta A_\nu, F_{\mu\nu}]
+ c_3 (\ep) {1 \over 2} D_\alpha D_\alpha \delta A_\mu\cr
&\quad + c_4 (\ep) {1 \over 2} (D_\mu D_\nu + D_\nu D_\mu) \delta
A_\nu + \tilde{c}(\ep) \{ \delta A_\nu, \tilde{F}_{\mu\nu} \}
\Big] \quad,\cr
}}
where we denote the $\ep$ dependence
of the counterterms explicitly.  For the derivation of the anomaly,
only the term proportional to $\tilde{c}(\ep)$ will play a role.

Before we apply the variational formula \evar\ to the chiral anomaly,
we determine the counterterm $\tilde{c}(\ep)$.  The second order variation
of the vacuum energy $W[A+\delta_1 A + \delta_2 A] - W[A+\delta_1 A]
- W[A+\delta_2 A] + W[A]$ can be calculated by using
eqn.~\edelvac\ with respect to $\delta_1 A$, and then by using
the variational formula \evar\ with respect to $\delta_2 A$.
The integrability of the vacuum energy demands that the second order
variation be symmetric with respect to $\delta_1 A$ and $\delta_2 A$.
This gives
\eqn\eintegrability{\eqalign{&
\int d^4 r \left( \delta_2 A_\mu^A (r) [\delta_1 A \cdot \S (\ep;A(r))]_\mu^A
- (1 \leftrightarrow 2) \right) \cr
&\quad =
\int d^4 r \left( \delta_1 A_\mu^A (r) \left( \Delta J_\mu^A (r;
A+\delta_2 A) - \Delta J_\mu^A (r;A)\right) -
(1 \leftrightarrow 2) \right)~.\cr}}
By calculating the right-hand side using the explicit form
\ecounter, we obtain
\eqn\etorsion{\eqalign{&
\delta_2 A_\mu^A [\delta_1 A \cdot \S (\ep;A)]_\mu^A
- (1 \leftrightarrow 2) \cr
&\quad =
{3 c \over 2} ~\ep_{\mu\nu\alpha\beta} \tr
(\delta_1 A_\mu \delta_2 A_\nu - \delta_2 A_\mu \delta_1 A_\nu)
F_{\alpha\beta} + (\rm{total}~\rm{derivatives})~.\cr}}
Substituting eqn.~\eS\ into the left-hand side, we obtain
\eqn\ectilde{\tilde{c}(\ep) = - {3 c \over 4}~,}
a constant independent of $\ep$.

Now we are ready to derive an expression for the chiral anomaly.
By applying the variational formula \evar\ twice, we can
compute $D_\mu \vev{\J_\mu^A}_A$ up to second order
in the external gauge field:
\eqn\esecond{\eqalign{&
D_\mu \vev{\J_\mu^A (0)}_A \simeq\cr
& - {1 \over 2} \int d^3 \Omega_\ep (r) A_\nu^B (r) \hat{r}_\mu
\int_{\scriptstyle r' \ge \eta\atop
\scriptstyle |r'-r| \ge \kappa} d^4 r'~A_\alpha^C (r')
\vev{J_\alpha^C (r') J_\nu^B (r) J_\mu^A (0)}_{A=0}^c \cr
& \qquad - {1 \over 2} \left(c_2(\ep) + {c_3(\ep) + c_4(\ep) \over 2} \right)
\tr (-i T^A) [A_\mu, \partial^2 A_\mu] \cr
& \qquad - {1 \over 2} \left( - c_2(\ep) + {c_3(\ep) + 3 c_4(\ep) \over
2}\right)
\tr (-i T^A) [A_\mu, \partial_\mu \partial_\nu A_\nu]\cr
& \qquad + {3 c \over 4}\ep_{\mu\nu\alpha\beta} \tr (-i T^A)
\{\partial_\mu A_\nu, \partial_\alpha A_\beta\} \cr
& - {1 \over 2} \Bigg(
f^{ABD} A_\mu^B (0) \int_{r'\ge \eta} d^4 r'~A_\alpha^C (r')
\vev{J_\alpha^C (r') J_\mu^D (0)}_{A=0}^c\cr
&\quad + \tr (-i T^A) \left[ A_\mu, {c_3 (\eta) \over 2}
\partial^2 A_\mu + c_4 (\eta) \partial_\mu \partial_\nu A_\nu \right]
\Bigg)~,\cr}}
where $\hat{r}_\mu \equiv r_\mu/r$, $d^3 \Omega_\ep (r)$ is the
solid angle element at radius $r=\ep$, and
the limits $\kappa \to 0$ and
$\eta \to 0$ must be taken before the limit $\ep \to 0$.

In deriving the above formula \esecond, we used the relation
\eqn\ederivative{\eqalign{&{\partial \over \partial r_\mu}
\int_{|r'-r| \ge \ep} d^4 r'~A_\nu^B (r')
\vev{J_\nu^B (r') J_\mu^A (r)} \cr
&\quad =
\int_{|r'-r| \ge \ep} d^4 r'~A_\nu^B (r')
\vev{J_\nu^B (r') \partial_\mu J_\mu^A (r)}\cr
&\qquad - \int d^3 \Omega_\ep (r'-r) ~A_\nu^B (r') {(r'-r)_\mu
\over \epsilon} \vev{J_\nu^B (r') J_\mu^A (r)}~,\cr}}
which is the euclidean analogue of the well-known
formula in Minkowski space for
the time derivative of the time-ordered product:
\eqn\eTproduct{\partial_\mu {\bf T} J^\mu (t,\vec{x}) J^\nu (0)
= {\bf T} \partial_\mu J^\mu (t,\vec{x}) J^\nu (0) +
\delta (t) [J^0 (0,\vec{x}),J^\nu (0)]~.}

We can simplify
the second order expression \esecond\ in two steps.  First
using the OPE
\eqn\eope{\eqalign{&\lim_{\ep \to 0}
\int d^3 \Omega_\ep (r) A_\nu^B (r) \hat{r}_\mu \vev{
J_\alpha^C (r') J_\nu^B (r) J_\mu^A (0)}_{A=0}^c\cr
& \quad = - f^{ABD} A_\mu^B (0) \vev{J_\alpha^C (r') J_\mu^D
(0)}_{A=0}^c~,\cr}}
which is the euclidean analogue of
the current-current commutation relation,
we can restrict the range of integration over $r'$
to $r_0 \ge r'$, where $r_0$ is an arbitrary finite radius.
This assures the locality of the anomaly, since $r_0$
can be as small as we want.  Second, we can take the
gauge potential at $r=0$ as vanishing:
\eqn\eAzero{A (r=0) = 0~.}
This is because those contributions from the double integral which are
proportional to $A(0)$ are canceled by the counterterms.  We can
understand this cancelation as a consequence of the covariance under
the external gauge transformations.  After these simplifications, we
obtain the main result of this note:
\eqn\eanomaly{\eqalign{
&D_\mu \vev{\J_\mu^A (0)}_A \cr
&\simeq
- {1 \over 2} \partial_\sigma A_\nu^B \partial_\tau A_\alpha^C \lim_{\ep \to 0}
\int d^3 \Omega_\ep (r) r_\sigma \hat{r}_\mu \cr
&\qquad \times \lim_{\eta, \kappa \to 0}
\int_{\scriptstyle r_0 \ge r' \ge \eta\atop
\scriptstyle |r'-r| \ge \kappa} d^4 r'~ r'_\tau \vev{J_\alpha^C (r')
J_\nu^B (r) J_\mu^A (0)}_{A=0}^c \cr
& \quad + {3 c \over 4} ~\ep_{\mu\nu\alpha\beta} \tr (-i T^A)
\{ \partial_\mu A_\nu, \partial_\alpha A_\beta \}~.\cr}}
It is important to note the contribution of the counterterm.  The
radius $r_0$ is arbitrary, and the locality of the anomaly is
manifest.  The above expression for the anomaly is valid not only for
the free theory but also for the interacting theories like QCD with
quarks.  We can appreciate the subtlety of the double limit by
noticing, as in \rwilson, that if we take the limit $\ep \to 0$ before
$\eta \to 0$, the entire double integral in eqn.~\esecond\ is canceled
by the single integral due to the OPE \eope.

By combining eqn.~\eanomaly\
with the expected covariant anomaly equation \ecovanomaly, we can
obtain the following formula that determines the coefficient $c$
in terms of a double integral\foot{This formula gives the
euclidean analogue of the formulas (7.19 -- 21) of ref.~\rwilson.}:
\eqn\ec{\eqalign{&\partial_\sigma A_\nu^B
\partial_\tau A_\alpha^C \lim_{\ep \to 0}
\int d^3 \Omega_\ep (r) r_\sigma \hat{r}_\mu \cr
&\qquad \times \lim_{\eta, \kappa \to 0}
\int_{\scriptstyle r_0 \ge r' \ge \eta\atop
\scriptstyle |r'-r| \ge \kappa} d^4 r'~ r'_\tau \vev{J_\alpha^C (r')
J_\nu^B (r) J_\mu^A (0)}_{A=0}^c \cr
&= - {3 c\over 2} ~\ep_{\sigma\nu\tau\alpha} \tr (-i T^A)
\{\partial_\sigma A_\nu , \partial_\tau A_\alpha\}~.\cr}}
Without the contribution of the counterterm
in eqn.~\eanomaly, we would get a wrong factor $- 3 c$
on the right-hand side.

Since QCD is asymptotic free,
we can evaluate the coefficient $c$ using the free fermion theory.
For the free right-handed currents, we obtain
\eqn\ejjj{\vev{J_\alpha^C (r') J_\nu^B (r) J_\mu^A (0)}_{free}^c
= {(r'-r)_\beta r_\gamma r'_\delta \over
r'^4 r^4 (r'-r)^4}~T_{\alpha\nu\mu,\beta\gamma\delta}^{CBA}~,}
where
\eqn\eT{T_{\alpha\nu\mu,\beta\gamma\delta}^{CBA} \equiv
{i \over 8 \pi^6} \left(
\tr T^C T^B T^A ~{\rm Sp}~{1 - \gamma_5 \over 2} \gamma_\alpha
\gamma_\beta \gamma_\nu \gamma_\gamma \gamma_\mu \gamma_\delta
- (B \leftrightarrow C, \alpha \leftrightarrow \nu, \gamma \leftrightarrow
\delta) \right)~.}
Using the formula
\eqn\eintegral{\eqalign{
&\lim_{\eta, \kappa \to 0}
\int_{\scriptstyle r_0 \ge r' \ge \eta\atop
\scriptstyle |r'-r| \ge \kappa} d^4 r'
{(r'-r)_\beta r'_\delta r'_\tau \over
r'^4 (r'-r)^4} \cr
&\quad = {\pi^2 \over r}
\left[ \left( {1 \over 4} - {r^2 \over 6 r_0^2} \right)
(\delta_{\tau\beta} \hat{r}_\delta + \delta_{\delta\beta} \hat{r}_\tau)
+ \left( - {1 \over 4} + {r^2 \over 12 r_0^2}\right) \delta_{\tau\delta}
\hat{r}_\beta - {1 \over 2} \hat{r}_\beta \hat{r}_\tau \hat{r}_\delta
\right]~,\cr}}
it is straightforward to find the well-known answer
\eqn\ec{c = {1 \over 24 \pi^2}~.}

In lieu of conclusions, the author wishes to mention that the present
work is a byproduct of his recent efforts \ref\rhid{H.~Sonoda, ``The
Energy-Momentum Tensor in Field Theory I, II,'' hep-th/9504113,
9509018} to reformulate quantum field theory in coordinate space with
the help of the first order variational formulas analogous to eqn.~\evar.

\listrefs
\bye